\documentclass[a4paper,11pt]{article}
\usepackage{graphicx}
\title{Linear-scaling quantum Monte Carlo with \\ non-orthogonal localized orbitals}
\author{D. Alf\`{e}$^{\star \ddag}$ and M. J. Gillan$^\star$ \\
$^\star$Physics and Astronomy Department, University College London, \\ Gower Street,
London WC1E 6BT, UK\\
$^\ddag$Department of Earth Sciences, University College London, \\
Gower Street, London WC1E 6BT, UK}

\begin{document}
\maketitle

\begin{abstract}
We have reformulated the quantum Monte Carlo (QMC) technique
so that a large part of the calculation scales linearly
with the number of atoms. The reformulation is related
to a recent alternative proposal
for achieving linear-scaling QMC, based on 
maximally localized Wannier orbitals (MLWO),
but has the advantage of greater simplicity.
The technique we propose draws on methods
recently developed for
linear-scaling density functional theory.
We report tests of the new technique
on the insulator MgO, and show that
its linear-scaling performance is somewhat
better than that achieved by the MLWO approach.
Implications for the application of QMC to
large complex systems are pointed out.
\end{abstract}

The quantum Monte Carlo (QMC) technique~\cite{foulkes01}
is becoming ever more important in the study of
condensed matter, with recent applications
including the reconstruction of
semiconductor surfaces~\cite{healy01},
the energetics of point defects in 
insulators~\cite{hood03}, optical
excitations in nanostructures~\cite{williamson02},
and the energetics of organic molecules~\cite{aspuru04}.
Although its demands on computer power are much
greater than those of widely used techniques
such as density functional theory (DFT),
its accuracy is also much greater for
most systems. With QMC now being applied to
large complex systems containing hundreds
of atoms, a major issue is the scaling of
the required computer effort with system
size. In other electronic-structure
techniques, including DFT, the locality
of quantum coherence~\cite{kohn96} suggests
that it should generally be possible
to achieve linear-scaling, or $O(N)$
operation, in which the computer effort
is proportional to the number of atoms $N$.
Very recently, a procedure has been
suggested~\cite{williamson01} for achieving
at least partial linear scaling for QMC,
based on the idea of ``maximally localized
Wannier functions''~\cite{marzari97}. The purpose of this
report is to propose and test a
simpler alternative method, which appears
to have important advantages.


The $O(N)$ techniques that have been developed
for tight binding (TB)~\cite{bowler97}, 
DFT~\cite{bowler02,soler02} and 
Hartree-Fock~\cite{challacombe99} calculations
all depend ultimately on the fact that the density
matrix $\rho ( {\bf r} , {\bf r}^\prime )$
associated with the single-electron orbitals
decays to zero as $| {\bf r} - {\bf r}^\prime | \rightarrow \infty$,
and the manner of this decay has been extensively 
studied (\cite{ismail99} and references therein).
Briefly, the decay is algebraic for metals and
exponential for insulators, with the decay rate
increasing with band gap, so that there is more
to be gained from $O(N)$ techniques for wide-gap
insulators. Equivalently, the extended orbitals
used in most conventional techniques can be linearly
combined to form localized orbitals, which again
decay exponentially in insulators. 
(For a review of early localized-orbital methods
in quantum chemistry, see Ref.~\cite{stoneham75}.) Orthogonal
Wannier functions are one form of localized orbitals,
but it has long been recognized that stronger localization
can be achieved by going to non-orthogonal orbitals,
as is done in many existing $O(N)$ TB, DFT 
and quantum-chemistry techniques.

In QMC, the trial many-body wavefunction
$\Psi_{\rm T} ( {\bf r}_1 , \ldots {\bf r}_N )$ consists of
a Slater determinant $D$ of single-electron
orbitals $\psi_n ( {\bf r}_i )$ multiplied by a parameterized Jastrow
correlation factor $J ( {\bf r}_1 , \ldots {\bf r}_N )$.
(In the pseudopotential-based QMC of interest here,
the $\psi_n ( {\bf r}_i )$ are commonly taken
from a plane-wave pseudopotential DFT
calculation.)
In variational Monte Carlo (VMC), $J$ is `optimized'
by varying its parameters so as to reduced the variance
of the `local energy' $\Psi_{\rm T}^{-1} \left( \hat{H} \Psi_{\rm T} \right)$,
where $\hat{H}$ is the many-electron Hamiltonian.
Since VMC by itself is not usually accurate enough, the
optimized $\Psi_{\rm T}$ produced by VMC is used in diffusion
Monte Carlo (DMC), which achieves the exact ground state
within the fixed nodal structure imposed by the Slater
determinant $D$. In conventional DMC, a large fraction of
the computer time goes into evaluating the single-electron
orbitals for all the electron positions ${\bf r}_i$ in each
of the replicas (QMC ``walkers''). For this part of
the calculations, the number of computer operations
required to perform each QMC step scales at least as $N^2$ ($M$
orbitals $\psi_n ( {\bf r}_i )$ for $N$ positions
${\bf r}_i$, with $M$ proportional to $N$), and the
scaling deriorates to $N^3$ if (as is often done) a plane-wave
basis set is used to represent the $\psi_n ( {\bf r}_i )$.
The memory requirement scales as $N^2$, and this also 
places important limits on the size of system that can
be treated.

As pointed out by Williamson {\em et al.}~\cite{williamson01}, the scaling for
the evaluation of the $\psi_n ( {\bf r}_i )$ can be
reduced from $N^3$ or $N^2$ to $N$ 
if the Slater determinant
$D$ is re-expressed in terms of localized orbitals, which
themselves are represented in terms of localized basis functions. The key
point here is that a determinant is changed by at most
a constant overall factor if arbitrary linear combinations
of its rows or columns are made. More precisely, 
if we construct orbitals $\phi_m ( {\bf r} )$
as linear combinations
\begin{equation}
\phi_m ( {\bf r} ) = \sum_{n=1}^M c_{m n} \psi_n ( {\bf r} ) \; ,
\; \; \; \; \; m = 1, \, 2, \, \ldots \, M
\end{equation}
of the original single-electron orbitals $\psi_n ( {\bf r} )$,
then the determinant ${\rm det} \, | \phi_m ( {\bf r}_i ) |$ is
given by:
\begin{equation}
{\rm det} \, | \phi_m ( {\bf r}_i ) | = {\rm det} \, | c_{m n} | \, . \,
{\rm det} \, | \psi_n ( {\bf r}_i ) | \; .
\end{equation}
This means that the determinants ${\rm det} \, | \phi_m ( {\bf r}_i ) |$
and ${\rm det} \, | \psi_n ( {\bf r}_i ) |$ are exactly the same functions
of the $M$ electronic positions $\{ {\bf r}_i \}$, apart from the constant
factor ${\rm det} \, | c_{m n} |$, so that, provided the
transformation matrix $|| c_{m n} ||$ is non-singular,
the orbitals $\phi_m ( {\bf r}_i )$ yield precisely
the the same total energy in QMC as the $\psi_n ( {\bf r} )$.
It is important to appreciate
that the linear combinations need not constitute
a unitary transformation, so that the 
orbitals $\phi_m ( {\bf r} )$ can be non-orthogonal.
The additional freedom gained by dispensing with
orthogonality can be exploited to improve the localization
of the orbitals.

These ideas lead to the following simple and general scheme.
Let $\psi_n ( {\bf r} )$ be any set of
$M$ orbitals, which in general extend over the
entire volume $\Omega$ of the cell containing
the atoms, and let there be a region $\omega$ of arbitrary shape
contained in $\Omega$. We wish to find the
linear combination $\phi ( {\bf r} ) =
\sum_{n=1}^M c_n \psi_n ( {\bf r} )$ such that
$\phi ( {\bf r} )$
is maximally localized in $\omega$. We interpret
this to mean that the $c_n$ are varied so as
to maximize the quantity $P$ defined by:
\begin{equation}
P = \int_\omega d {\bf r} \, | \phi ( {\bf r} ) |^2 \left/
\int_\Omega d {\bf r} \, | \phi ( {\bf r} ) |^2 \right. \; ,
\end{equation}
where the integrals go over the regions $\omega$
and $\Omega$ respectively. We refer to
$P$ as the localization weight,
and note that by definition $0 \le P \le 1$.
Now $P$ can be re-expressed as:
\begin{equation}
P = \sum_{m, n} c_m^\star A_{m n}^\omega c_n \left/
\sum_{m, n} c_m^\star A_{m n}^\Omega c_n \right. \; ,
\end{equation}
where:
\begin{equation}
A_{m n}^\omega = \int_\omega d {\bf r} \, \psi_m^\star \psi_n \; ,
\; \; \; \; \; A_{m n}^\Omega = \int_\Omega d {\bf r} \,
\psi_m^\star \psi_n \; .
\end{equation}
Clearly $P$ takes its maximum value when the $c_n$
are the components of the eigenvector of the
generalized eigenvalue equation:
\begin{equation}
\sum_n A_{m n}^\omega c_n = \lambda_\alpha \sum_n A_{m n}^\Omega c_n 
\end{equation}
associated with the largest eigenvalue $\lambda_1$,
and this maximum value of $P$ is equal to $\lambda_1$.
More generally, the $s$ most localized linear combinations
in $\omega$ are associated with the $s$ largest eigenvalues
$\lambda_\alpha$ ($\alpha = 1, \ldots s$),
where the $\lambda_\alpha$ are ordered in descending sequence.

The idea is now to use this scheme to produce $M$ localized orbitals
$\phi_m ( {\bf r} )$, which are used instead of the
$M$ extended orbitals $\psi_n ( {\bf r} )$ in the determinant $D$. 
We do this by choosing a number $\nu_{\rm reg}$ of localization
regions, and taking the $p_{\rm loc}$ most localized orbitals
in each of these regions, so that $M = \nu_{\rm reg} p_{\rm loc}$
is equal to the number of orbitals $\psi_n ( {\bf r} )$. If no
approximations are made, the results thus obtained in both
VMC and DMC will be identical to those obtained with the $\psi_n$.
But now we can obtain $O(N)$ scaling by truncating the
$\phi_m ( {\bf r} )$ so that they are exactly zero
outside their localization regions. Provided the
$\phi_m ( {\bf r} )$ have almost all their weight
inside their regions, the only effect of this truncation
will be to introduce a slight shift of the nodal surfaces,
so that DMC results will come out essentially identical
to those obtained with the conventional $O ( N^2 )$ scheme.
This gives $O(N)$ scaling because it makes the determinant
$| \phi_m ( {\bf r}_i ) |$ sparse. Specifically,
for a given electron position ${\bf r}_i$,
the number of orbitals $\phi_m$ for which
$\phi_m ( {\bf r}_i )$ is non-zero is no longer
the total number of orbitals, but is proportional
to the number of localization regions that contain
${\bf r}_i$. This is $O(1)$, rather than $O(N)$,
and the number of operations needed for orbital
evaluation is reduced by a factor equal to the ratio
of the localization volume to the volume of
the whole system.

To implement this method, we need to consider carefully
the basis used for representing the $\phi_m ( {\bf r} )$
and the $\psi_n ( {\bf r} )$. This basis must be large enough
and flexible enough to represent them very accurately, while
at the same time representing the truncated $\phi_m ( {\bf r} )$
everywhere without introducing computationally troublesome
discontinuities. To achieve linear scaling,
it is also vital that the basis functions
be localized. Fortunately, this problem of representing
localized orbitals has already been extensively
studied within $O(N)$ 
DFT~\cite{mostofi02,gan01,fattebert00,hernandez96}. 
We adopt here the B-spline
(also called `blip-function') representation used
in the {\sc conquest} $O(N)$ DFT code~\cite{bowler02,hernandez96}. The blip functions
consist of localized cubic splines centred on the points
of a regular grid, each function being non-zero only inside a region
extending two grid spacings in each direction from its centre.
Details of the blip representation, its close relationship
with the plane-wave basis, and its efficiency in representing
both extended and localized functions have been described
elsewhere~\cite{hernandez97}.

We have tested our method on the prototypical oxide material MgO in
the rock-salt structure at ambient pressure. Because of its large band
gap (experimental $E_{\rm g} = 7.7$~eV), this kind of material should
be particularly suited to $O(N)$ methods.  The QMC calculations were
performed using the appropriately modified {\sc casino}
code~\cite{needs04} on a supercell of 64 atoms, with extended orbitals
$\psi_n ( {\bf r} )$ at the $\Gamma$-point obtained from a DFT
plane-wave calculation in the local-density approximation.
Hartree-Fock pseudopotentials were used.  Since MgO is a highly ionic
material, the valence electrons are associated almost entirely with
the anions, so that it is natural to take localization regions
associated with the O ions. We take four localized orbitals $\phi_m (
{\bf r} )$ for each O ion. The simplest procedure is to localize all
four orbitals in the same region centred on the O site, this region
being either spherical or cubic. We then expect to find one most
localized orbital, corresponding roughly to an O 2s state, and three
next most localized orbitals, corresponding roughly to the three 2p
states. However, another way of thinking about localization is to seek
four equivalent hybrid sp$^3$-like orbitals. If we do this, it is
natural to take four separate localization regions for each O, one for
each sp$^3$-like orbital. With this alternative procedure, to preserve
the symmetry between the four orbitals, we take the centres of the
regions to be displaced by a distance $d$ from the O site along four
tetrahedral directions, and we keep only the single most localized
orbital in each region. The distance $d$ can be chosen to optimize
the localization weight.

In order for our method to succeed, the localized orbitals $\phi_m (
{\bf r} )$ must be strongly localized (localization weight $P$ very
close to unity) in regions having reasonably small cut-off radii.
Fig.~1 compares the calculated localization weights $P$ (the quantity
plotted is actually $Q \equiv 1 - P$) with those of the maximally
localized Wannier functions produced by the scheme of
Ref.~\cite{williamson01} for both cubic and spherical regions.  We
note that the results shown for our method were obtained with
localization regions centred on O sites, while the Wannier results
were obtained with displaced centres.  In spite of this, our $\phi_m (
{\bf r} )$ are remarkably strongly localized, even for cut-off radii
as small as 6~a.u.  The decay of $Q$ to zero is much more rapid for
non-orthogonal orbitals, as expected. Also expected is the more rapid
decay in both methods if cubic, rather than spherical regions are
used. We have done tests also on the localization weight in our method
when the localization centres are displaced, and we find that a
displaced distance of $d=0.66$ a.u. works well. The main effect of
this is to reduce $Q$ for all four orbitals to roughly the value for
the s-like orbital in Fig. 1.

for the p-like orbitals to essentially the value
obtained for the s-like orbitals.

Given the excellent localization, 
the key question is now the convergence of the VMC and
DMC total energies obtained in our $O(N)$ scheme
towards the `exact' values obtained with extended
orbitals. We show in Fig.~2 the total energy per atom
obtained in VMC from calculations using spherical
and cubic cut-offs and with localization centres
on O sites and on displaced sites. As expected,
convergence is somewhat more rapid with cubic regions,
though the difference is not large. There is
a considerable improvement with displaced localization
centres. With this procedure, the total energy
agrees with the `exact' value already within
$\sim 10$~meV/atom when the localization radius
is 6~a.u., and even for a radius of 4.5~a.u. the
agreement is only slightly worse. For comparison,
we show results obtained with orthogonal Wannier 
orbitals, these orbitals being on displaced centres. 
We note that the total energy converges considerably
less rapidly. With spherical regions, the energy
still differs from the `exact' value by at least
20~meV even with a radius of 7.5~a.u., which is almost
half the cell length. Timings on our scheme
show, as expected, that the time for evaluation
of non-zero orbitals $\phi_m ( {\bf r}_i )$ is
proportional to the ratio of
the localization volume to the volume
of the whole cell. For example, if we go
from a cubic localization region having a radius
of 7.8~a.u. to a spherical region of radius 6~a.u.,
the time for orbital evaluation decreases
by a factor of roughly $( 4 \pi 6^3 / 3 ) / ( 2 \times 7.8 )^3
\simeq 0.25$.

We have also done tests on the performance of our
linear-scaling method for DMC. We show in Fig.~3 the
total energy per atom for MgO from DMC calculations 
using a localization region centred on the O sites. As
with the VMC tests, the energy appears to be converged
within $\sim 10$~meV/atom for a cut-off radius of 6~a.u.
As in the case of VMC, the time for orbital evaluation
is proportional to the localization volume.

We have therefore demonstrated that a precision
of 10~meV/atom, which is considerably better than
is needed for most purposes, is achieved with
remarkably small localization regions. Even though
our tests have all been done on a single system size,
we can be sure that $O(N)$ scaling is achieved,
because the reduction in the number of operations
is proportional to the ratio of the localization
volume to the volume of the whole system.

An additional and important gain from the type of $O(N)$ QMC
proposed here is the large factor by which the
memory requirement is reduced. This factor is also
roughly the ratio between the volume of the localization 
region and the volume of the whole system. With the
rather small MgO system of 64 ions used for the present
tests, and a spherical localization cut-off of 6~a.u.,
the memory is reduced by a factor of 4. But QMC
calculations on extended matter frequently needs systems
of up to 10 times this size, so that the memory
requirement will often be reduced by well over
an order of magnitude. Although the number of operations
needed for orbital evaluation is also reduced
by this factor, at present the overall gain
is less dramatic, since other parts of the QMC
calculation (Ewald calculation of Coulomb energy,
manipulation of determinants, etc) may account for
up to 50~\% of the time. Nevertheless,
this still means a halving of the computation
time for large systems, even if these other parts
of QMC are left as they are. However, as pointed
out earlier~\cite{williamson01}, $O(N)$ operation
should be achievable for these other parts too.

We also note that the techniques we have described
may be the key to constructing QMC `embedding'
methods, in which QMC is used to treat
a spatially localized part of an extended
system (e.g. a defect or a surface),
while the rest of the system is treated
either by using a matched technique of
lower precision, or by invoking $O(N)$
QMC information on the bulk. The close
relationship between the $O(N)$ problem
and the embedding problem has been
pointed out elsewhere~\cite{bowler02a}
in the context of tight-binding and DFT
calculations.

The oxide system we have used for the present
practical tests has the special feature of
a large band gap. It goes without saying
that useful $O(N)$ operation will not
be so easily obtained for metals and
narrow-gap semiconductors. However, large
areas of materials physics and chemistry,
as well as bio-materials and aqueous systems,
involved wide-gap materials, so we expect 
$O(N)$ QMC to be very widely applicable. 
In oxide science, problems where the
new method should be immediately applicable
include the energetics of bulk defects, perfect and
defective surfaces and the adsorption
of molecules at these surfaces, where
the deficiencies and limitations of standard DFT
methods are well known. We are currently preparing
to attempt such calculations on MgO systems.

In conclusion, we have proposed and tested a new technique
for achieving linear scaling in one of the most
demanding parts of QMC calculations. In addition
to being simpler and more robust than an earlier
technique, it appears also to be more efficient.
The new technique already makes it possible to treat
large systems that would be out of reach of
conventional QMC methods. Research areas where
the technique is immediately applicable include
defects and surfaces of oxide materials, and molecular
processes on these surfaces.
\bigskip

\noindent
DA acknowledges support from the Royal Society,
and also thanks the Leverhulme Trust and the CNR for
support. The authors are indebted to
R.J.~Needs, M.D.~Towler and N.D.~Drummond for advice and technical
support in the use of the {\sc casino} code,
and for providing us with the pseudopotentials
used in this work.

\newpage

\begin{figure}[htbp]
\centering
\includegraphics[clip,width=0.75\textwidth,angle=270]{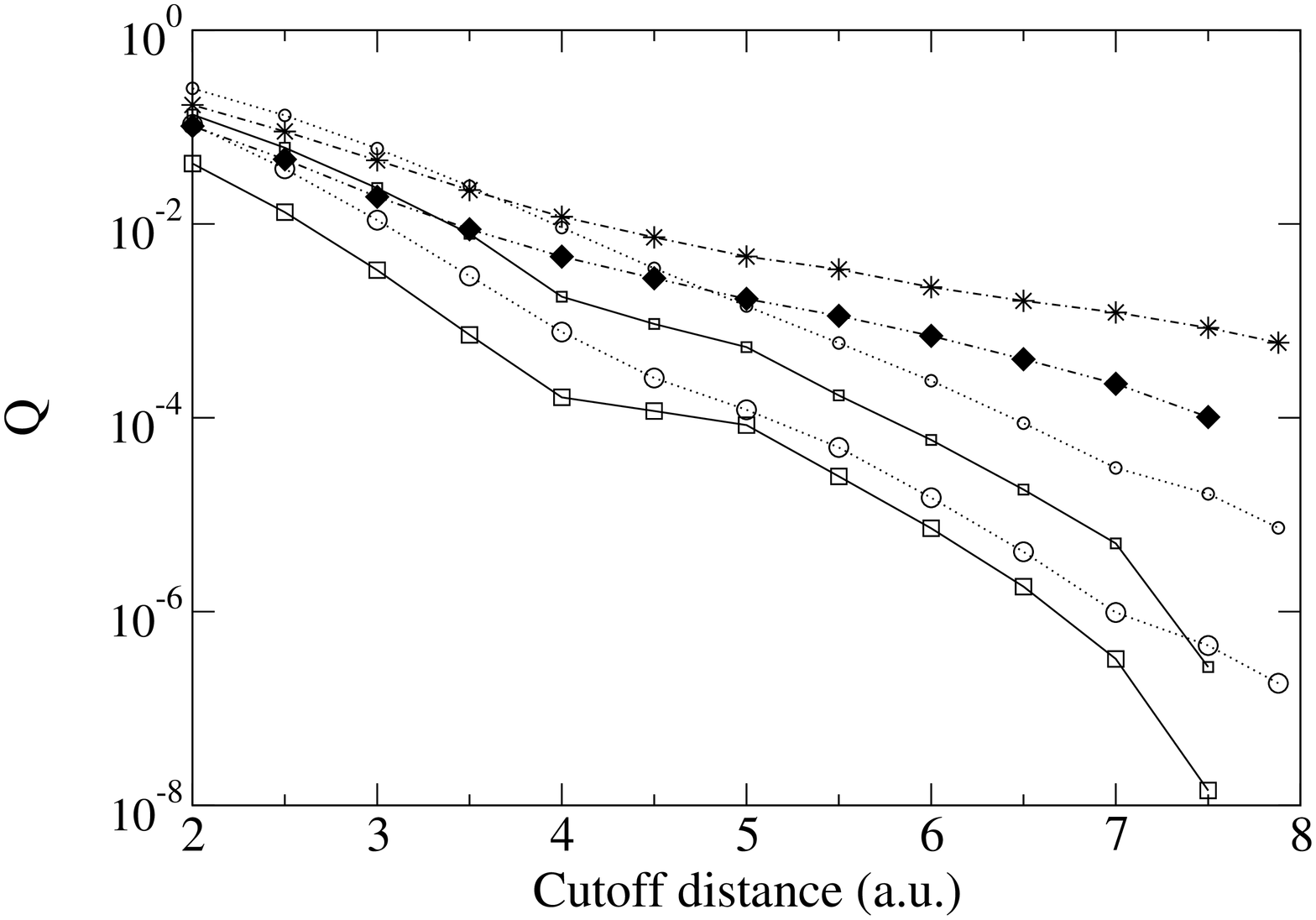}
\caption{Dependence of localization weight on cut-off distance
for bulk MgO. Quantity plotted is $Q \equiv 1 - P$, where $P$
is localization weight defined by eqn~(3). Large and small
open circles: spherical cut-off for s-like and p-like
localized orbitals; large and small open squares:
cubic cut-off for s-like and p-like localized orbitals.
Solid diamonds and stars: cubic and spherical cut-off
for maximally localized Wannier orbitals.}
\label{fig:loc_weight}
\end{figure}

\newpage

\begin{figure}[htbp]
\centering
\includegraphics[clip,width=0.75\textwidth,angle=270]{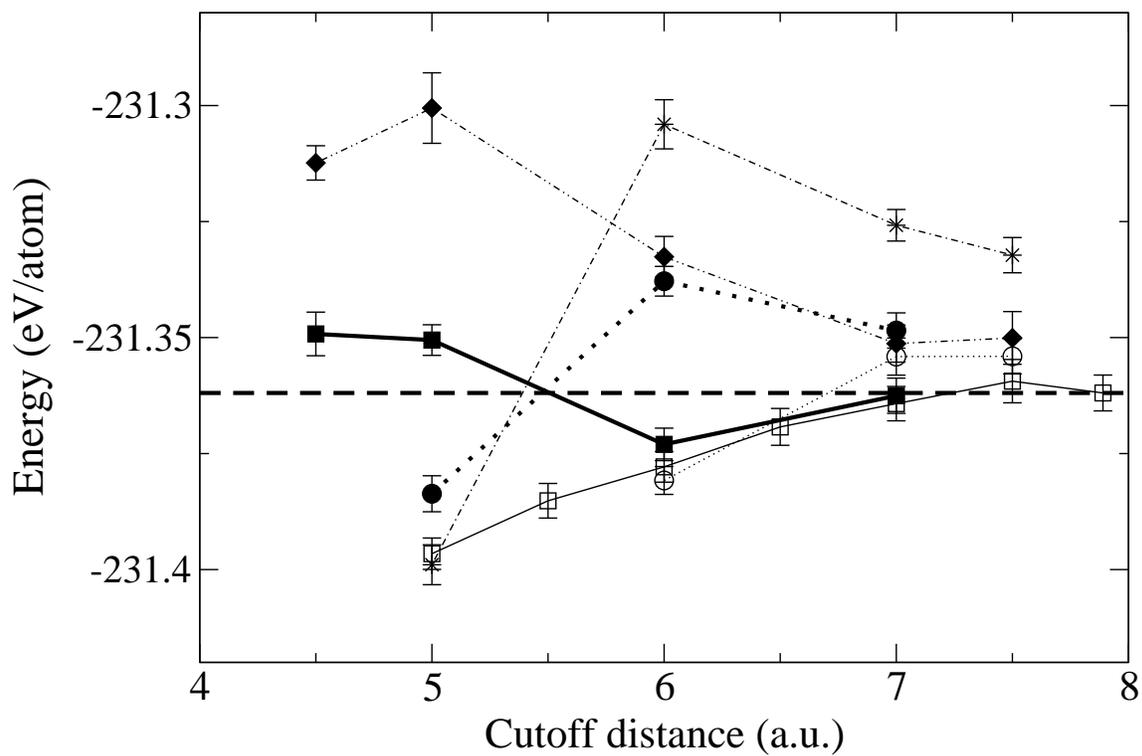}
\caption{Convergence of linear-scaling VMC total energy per
atom to value obtained with extended orbitals for bulk MgO.
Open circles and squares: present method with spherical
and cubic cut-offs, for localized orbitals centred on O sites.
Filled circles and squares: spherical and cubic cut-offs,
for displaced localized orbitals. Filled diamonds and
stars: maximally localized Wannier orbitals with cubic
and spherical cut-offs. Horizontal dashed line shows total
energy/atom obtained with extended orbitals.}
\label{fig:VMC}
\end{figure}

\newpage

\begin{figure}[htbp]
\centering
\includegraphics[clip,width=0.75\textwidth,angle=270]{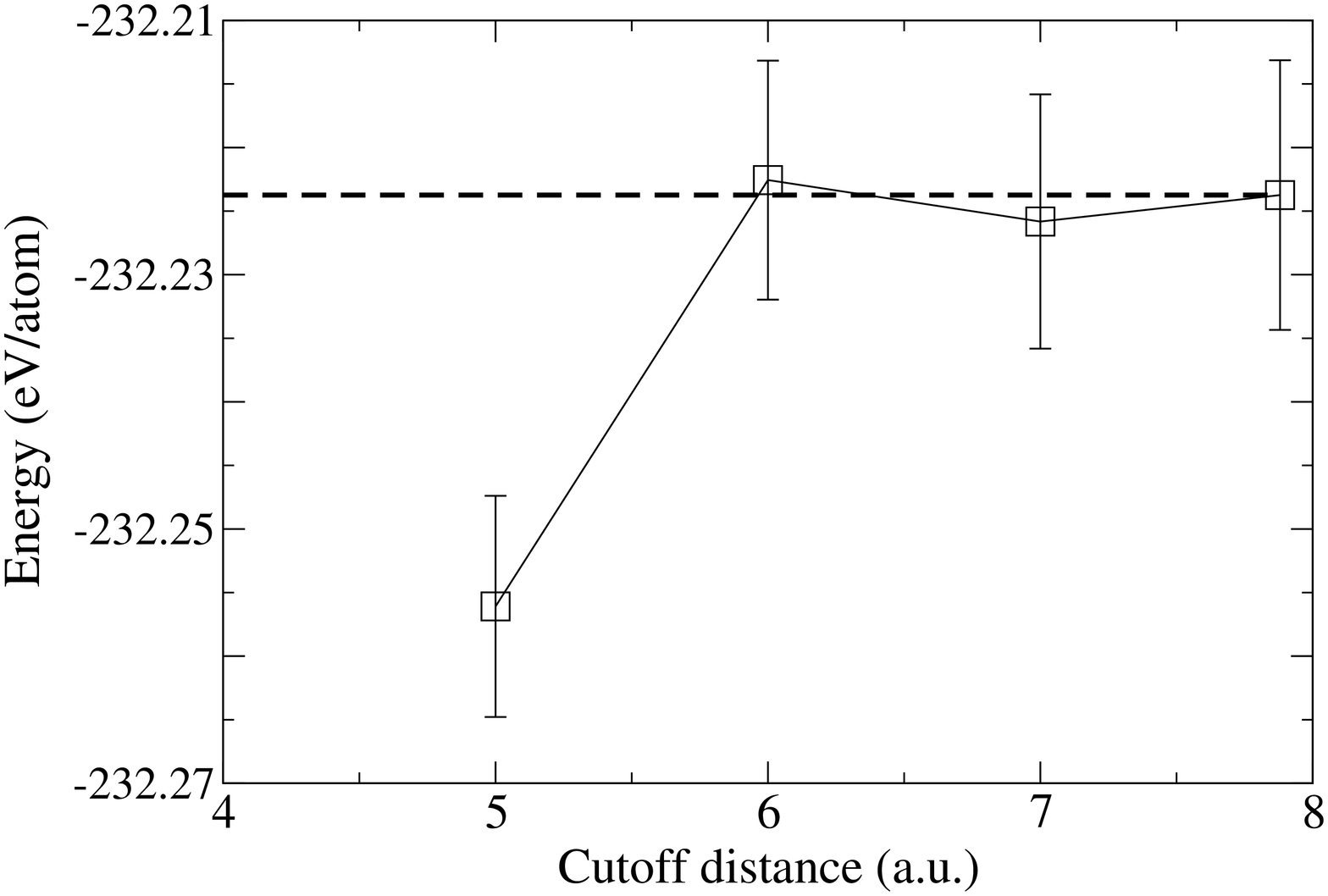}
\caption{Convergence of linear-scaling DMC total energy
per atom to value obtained with extended orbitals for bulk MgO.
Open squares with statistical error bars: present with
cubic cut-off. Dashed line shows result with extended orbitals.}
\label{fig:DMC}
\end{figure}


\begin{thebibliography}{99}

\bibitem{foulkes01}
Foulkes~W~M~C, Mitas~L, Needs~R~J and Rajagopal~G
2001 {\em Rev. Mod. Phys.} {\bf 73} 33

\bibitem{healy01}
Healy~S~B, Filippi~C, Kratzer~P, Penev~E and Scheffler~M 2001
{\em Phys. Rev. Lett.} {\bf 87} 016105

\bibitem{hood03}
Hood~R~Q, Kent~P~R~C, Needs~R~J and Briddon~P~R
2003 {\em Phys. Rev. Lett.} {\bf 91} 076403

\bibitem{williamson02}
Williamson~A~J, Grossman~J~C, Hood~R~Q, Puzder~A and Galli~G
2002 {\em Phys. Rev. Lett.} {\bf 89} 196803

\bibitem{aspuru04}
Aspuru-Guzik~A, El Akramine~O, Grossman~J~C and Lester~W~A
2004 {\em J. Chem. Phys.} {\bf 120} 3049

\bibitem{kohn96}
Kohn~W 1996 {\em Phys. Rev. Lett.} {\bf 76} 3168

\bibitem{williamson01}
Williamson~A~J, Hood~R~Q and Grossman~J~C 2001 {\em Phys. Rev. Lett.}
{\bf 87} 246406

\bibitem{marzari97}
Marzari~N and Vanderbilt~D 1997 {\em Phys. Rev. B} {\bf 56} 12847

\bibitem{bowler97}
Bowler~D~R, Aoki~M, Goringe~C~M, Horsfield~A~P and
Pettifor~D~G 1997 {\em Modell. Simul. Mater. Sci. Eng.}
{\bf 5} 199

\bibitem{bowler02}
Bowler~D~R, Miyazaki~T and Gillan~M~J 2002 {\em J. Phys.: Condens. Matter}
{\bf 14} 2781

\bibitem{soler02}
Soler~J~M, Artacho~E, Gale~J~D, Garcia~A, Junquera~J,
Ordejon~P and Sanchez-Portal~D 2002 {\em J. Phys.: Condens. Matter}
{\bf 14} 2745

\bibitem{challacombe99}
Challacombe~M 1999 {\em J. Chem. Phys.} {\bf 110} 2332

\bibitem{ismail99}
Ismail-Beigi~S and Arias~T~A 1999 {\em Phys. Rev. Lett.} {\bf 82} 2127

\bibitem{stoneham75}
Stoneham~A~M 1975 {\em Theory of Defects in Solids},
Oxford University Press, Oxford, Sec.~7.5

\bibitem{mostofi02}
Mostofi~A~A, Skylaris~C~K, Haynes~P~D and Payne~M~C 2002
{\em Comput. Phys. Commun.} {\bf 147} 788

\bibitem{gan01}
Gan~C~K, Haynes~P~D and Payne~M~C 2001 {\em Phys. Rev. B} {\bf 63} 205109

\bibitem{fattebert00}
Fattebert~J~L and Bernholc~J 2000 {\em Phys. Rev. B} {\bf 62} 1713

\bibitem{hernandez96}
Hern\'{a}ndez~E, Gillan~M~J and Goringe~C~M 1996 {\em Phys. Rev. B}
{\bf 53} 7147

\bibitem{hernandez97}
Hern\'{a}ndez~E, Gillan~M~J and Goringe~C~M 1997 {\em Phys. Rev. B}
{\bf 55} 13485

\bibitem{needs04}
Needs~R~J, Towler~M~D, Drummond~N~D and Kent~P~R~C 2004
`{\sc casino} Version 1.7 User Manual', University of Cambridge, 
Cambridge

\bibitem{bowler02a}
Bowler~D~R and Gillan~M~J 2002 {\em Chem. Phys. Lett.}
{\bf 355} 306

\end{thebibliography}
\end{document}